\documentclass[twocolumn,
                showpacs,
                nofootinbib,
                nobibnotes,
                aps,
                superscriptaddress,
                amssymb,
                amsmath,
                floatfix,
                reprint,
                prx,
                notitlepage,
                showkeys,
                10pt,
		longbibliography]{revtex4-1}
                
\usepackage{amssymb}
\usepackage{amsmath}
\usepackage{cprotect}
\usepackage{graphicx}
\usepackage{dcolumn}
\usepackage{url}
\usepackage[colorlinks=true,breaklinks=true,allcolors=blue]{hyperref}
\usepackage{tikz}
\usepackage{dsfont}
\usepackage{epsfig}
\usepackage{color}
\usepackage[ruled]{algorithm2e}
\usepackage{bbold}
\usepackage[binary-units=true]{siunitx}
\usepackage{glossaries}
\usepackage{tikz}
\usepackage{bm}
\usepackage{hhline}
\usepackage{spverbatim}
\usepackage{mhchem}
\usepackage{siunitx}

\makeatletter
\let\cat@comma@active\@empty
\makeatother

\newcommand{\Eq}[1]{Eq.\,\eqref{eq:#1}}
\newcommand{\Fig}[1]{Fig.~\ref{fig:#1}}

\newcommand{\App}[1]{App.~\ref{app:#1}}
\newcommand{\Sect}[1]{Sect.~\ref{sect:#1}}

\newcommand{\ket}[1]{|#1\rangle}
\newcommand{\bra}[1]{\langle#1|}
\newcommand{\SvN}[1]{S_{1}\left(#1\right)}
\newcommand{\pc}{p_{\mathrm{c}}}

\begin{document}
\title{Simulating a measurement-induced phase transition for trapped ion circuits}

\author{Stefanie Czischek}
\email{sczischek@uwaterloo.ca}
\affiliation{Department of Physics and Astronomy, University of Waterloo, Ontario, N2L 3G1, Canada}

\author{Giacomo Torlai}
\affiliation{AWS Center for Quantum Computing, Pasadena, CA, 91125, USA}

\author{Sayonee Ray}
\affiliation{Department of Physics and Astronomy, University of Waterloo, Ontario, N2L 3G1, Canada}
\affiliation{1QB Information Technologies (1QBit), Vancouver, British Columbia, Canada}

\author{Rajibul Islam}
\affiliation{Department of Physics and Astronomy, University of Waterloo, Ontario, N2L 3G1, Canada}
\affiliation{Institute for Quantum Computing, University of Waterloo, Ontario, N2L 3G1, Canada}

\author{Roger G. Melko}
\affiliation{Department of Physics and Astronomy, University of Waterloo, Ontario, N2L 3G1, Canada}
\affiliation{Perimeter Institute for Theoretical Physics, Waterloo, Ontario, N2L 2Y5, Canada}

\date{\today}

\begin{abstract}

The rise of programmable quantum devices has motivated the exploration of circuit models which could realize novel physics.
A promising candidate is a class of hybrid circuits, where entangling unitary dynamics compete with disentangling measurements. 
Novel phase transitions between different entanglement regimes have been identified in their dynamical states,
with universal properties hinting at unexplored critical phenomena.
Trapped ion hardware is a leading contender for the experimental realization of such physics, which requires not only traditional
two-qubit entangling gates, but also a constant rate of local measurements accurately addressed throughout the circuit. 
Recent progress in engineering high-precision optical addressing of individual ions makes preparing a constant rate of measurements throughout a unitary circuit feasible.
Using tensor network simulations, we show that the resulting class of hybrid circuits, prepared with native gates,
exhibits a volume-law to area-law transition in the entanglement entropy.  This displays universal hallmarks of a measurement-induced phase transition.
Our simulations are able to characterize the critical exponents using circuit sizes with tens of qubits and thousands of gates.  
We argue that this transition should be robust against additional sources of experimental noise expected in modern trapped ion hardware,
and will rather be limited by statistical requirements on post-selection.  Our work highlights the powerful role that tensor network
simulations can play in advancing the theoretical and experimental frontiers of critical phenomena.
\end{abstract}  

\maketitle

\section{Introduction}
The study of quantum criticality is experiencing a sudden transformation originating in the idea of measurement-induced phase transitions (MPTs)
~\cite{Li2018,Skinner2019,Chan2019}.
These exist in many-body qubit wavefunctions, where entangling unitary dynamics compete with non-unitary measurements
that act to disentangle the state.
Such phase transitions can be driven by a finite rate of measurement throughout the circuit, or by the
choice of measurement bases~\cite{Sang2020,Lavasani2021} or the strength of weak measurements~\cite{Szyniszewski2019,Bao2020,Szyniszewski2020,Doggen2021}.
Numerical studies of a variety of random quantum circuits have demonstrated that such phase transitions share broad features of conventional conformal 
field theories (CFTs)~\cite{Zabalo2020,Li2019,Gullans2020,Li2020,Skinner2019,Sang2020,Jian2020}, with some quantities depending on details of the circuit,
e.g., the interaction range between the qubits~\cite{Block2021,Minato2021}.
The observation of conventional CFT features, such as a critical scaling in the entanglement entropy characterized by universal exponents, was surprising and distinguishes the observed phase transition from a classical crossover.
While special cases have been related to well-known statistical or spin models~\cite{Jian2020,Bao2020,Fan2020,Li2021b}, percolation theories~\cite{Sang2020b,Shtanko2020},
or minimal-cut pictures~\cite{Nahum2021}, a general comprehensive theoretical framework is still absent and no rigorous connection to CFT has yet
been made.
This leaves open the prospect that MPTs may be offering a glimpse at a deeper, as-yet undiscovered set of critical points, and possibly a new class
of universal phenomena.
Further observations of MPTs in higher-dimensional systems~\cite{Turkeshi2020} or 
as purification phase transitions~\cite{Gullans2020b}, and their relation to quantum error correction~\cite{Choi2020,Li2021}
demonstrate the broad consequences such critical points can have in many fields of physics.

MPTs are known to exist in models, and significant debates exist as to whether they can be
realized in real experimental systems.
The most immediate candidate for engineering MPTs is a highly-controlled qubit system 
capable of programming a {\it hybrid} quantum circuit~\cite{Li2018,Skinner2019,Chan2019}.  
These circuits are composed of two ingredients: entangling unitary dynamics, and disentangling non-unitary measurements.
In most current experimental quantum devices, significant effort is being expended to produce high-quality two-qubit gates suitable for producing entanglement.
At sufficiently large circuit depths, these can induce a {\it volume law} entanglement scaling in the underlying quantum state.
Commonly-studied circuits to generate volume-law wavefunctions are built out of two-qubit unitaries drawn from the Clifford group or the random
Haar measure~\cite{Huang2021,Nahum2017,Nahum2018,Keyserlingk2018}.

The second ingredient typically used to induce MPTs in hybrid circuits are repeated local measurements 
applied at a constant rate $p$ throughout the circuit.  These measurements act to disentangle the wavefunction, 
and for large $p$ will reduce the entanglement to an {\it area law} scaling.
It is the intermediate regime of small but finite $p$ where conditions are ripe for MPTs to exist.  

The experimental realization of a quantum system harboring an MPT requires accurate programmable manipulation of individual qubits.
Experimental platforms based on laser-cooled trapped ions achieve such a high degree of control, where highly-pure
quantum states of tens of ions are routinely prepared~\cite{Monroe2021,Debnath2016}.
In a trapped ion computer, the entangling dynamics of a random quantum circuit can be achieved with a combination of native two-qubit entangling M{\o}lmer-S{\o}rensen (MS)
gates and random single-qubit rotations.
In addition, recent experiments employing programmable and high-precision single-ion optical addressing technologies~\cite{Crain2014, Shih2021} demonstrate the feasibility of a constant rate of measurements throughout the circuit.

In this article, we propose a class of hybrid circuit which is natural to implement using this set of native trapped ion gates.
We focus on non-Clifford circuits and study the behavior of the entanglement scaling using state-of-the-art tensor network algorithms,
capable of simulating tens of qubits acted on by thousands of gates.
We find a class of circuit that exhibits a phase transition in entanglement under a number of plausible experimental conditions.
We investigate the critical point, which displays universal hallmarks of an MPT, and calculate the critical measurement density
and the universal scaling exponent of the correlation length.
We argue that this transition should be robust against sources of experimental noise expected in modern trapped ion hardware, such as
incoherent gate errors, state preparation and measurement (SPAM) errors, and optical cross-talk between the target and neighboring ions.
We further demonstrate that bringing the qubit into an outcome-independent state after each measurement does not influence the MPT.
This enables an efficient experimental protocol and mitigates potential leakage of ions into atomic states outside the qubit Hilbert space during projective measurements \cite{Olmschenk2007}.
Beyond this robustness, near-term experiments will likely be limited by statistical requirements on post-selection~\cite{Nahum2021,Gullans2020,Bao2020}.
Due to the inherent random behavior of measurement outcomes, replicating the exact state after every run of the circuit becomes 
the main experimental challenge, especially for large system sizes and higher measurement rates. 
We conclude with a discussion of some recent proposals which may be able to bypass this bottleneck, 
such as detecting the MPT in a local order parameter via coupling a reference qubit to the system~\cite{Gullans2020,Lavasani2021,Bao2020,Noel2021},
or considering unitary evolutions in space-time duals~\cite{Ippoliti2021}.

\section{Hybrid circuit design}

We consider $N$ trapped atomic ions, each with two internal energy levels behaving as an effective qubit system.
In current experimental hardware, these two qubit states have been shown to be long-lived and highly coherent~\cite{Wang2021},
with accurate initialization through optical pumping and qubit readout via fluorescence imaging.
The qubits are manipulated coherently~\cite{Monroe2021} using external laser fields, capable of applying phase-sensitive single- and two-qubit quantum gates mediated by collective vibrational (phonon) modes.

\begin{figure}
	\includegraphics[width=\linewidth]{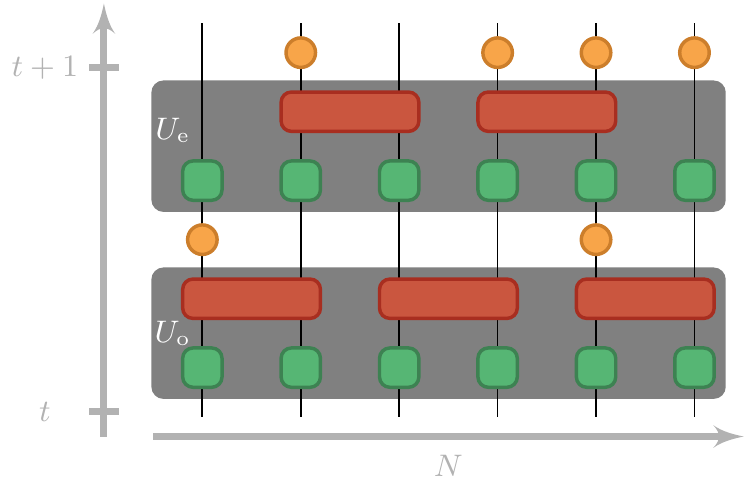}
	\caption{
	One time cycle realization of a quantum circuit setup to probe measurement-induced phase transitions, with vertical lines 
	indicating the $N=6$ evolved qubits.
	Squares denote single-qubit unitary rotation gates $R\left(\theta,\varphi\right)$, rectangles correspond to 
	two-qubit M{\o}lmer-S{\o}rensen gates $M\left(\Theta\right)$,
	and circles show single-qubit projective measurements in the $Z$-basis, which are injected uniformly in space and time
        at a rate $p$.
	The time cycle consists of an even (upper, $U_{\mathrm{e}}$) and an odd (lower, $U_{\mathrm{o}}$) unitary layer, 
	indicated in gray, each followed by a measurement layer. 
	In the even unitary layer the outer two qubits are not affected by the M{\o}lmer-S{\o}rensen gates due to open boundary 
	conditions.
	}
	\label{fig:1}
\end{figure}

We focus on modeling the dynamics generated by a hybrid circuit consisting of a mixture of entangling gates and disentangling measurements,
acting on a pure initial state $\ket{\Psi_0}=\ket{0}^{\otimes N}$.
To begin we examine randomized quantum gates that are native to the hardware, for their ability to induce strong entanglement in the wavefunction.
In trapped ion hardware, the native entangling gate--the MS gate--is equivalent to a rotation around the $XX$-axis by an angle $\Theta$,
with unitary propagator acting on qubits $j$ and $k$~\cite{Pogorelov2021}:
\begin{align}
	&M_{j,k}\left(\Theta\right)=
	\left[\begin{matrix}
	\mathrm{cos}\left(\Theta\right) & 0 & 0 & -i\mathrm{sin}\left(\Theta\right) \\
	0 & \mathrm{cos}\left(\Theta\right) & -i\mathrm{sin}\left(\Theta\right) & 0 \\
	0 & -i\mathrm{sin}\left(\Theta\right) & \mathrm{cos}\left(\Theta\right) & 0 \\
	-i\mathrm{sin}\left(\Theta\right) & 0 & 0 & \mathrm{cos}\left(\Theta\right)
	\end{matrix}\right].
\end{align}
There are multiple ways of introducing randomness in the circuit, such as the angles of each MS gate, the rotation axes, or the gate
connectivity, for example.
Here we study quantum circuits constructed in a brick-layer fashion (as illustrated in \Fig{1}), with each layer consisting of a set of MS
gates between neighboring qubits (alternating the bonds between layers), supplemented by a layer of random single-qubit rotations.
Those circuits have a straight-forward implementation in trapped ion experiments.
For the MS gates, we fix the angle to $\Theta=\pi/4$ uniformly throughout the circuit, generating maximal entanglement between the pairs 
at each application.
Note that we restrict connectivity to nearest neighbors only, despite the capability of trapped ion hardware to execute long-range gates.
As will become clear in the next section, this choice allows us to leverage powerful tensor-network algorithms for simulating the circuit dynamics.

The native random single-qubit gates for trapped ion hardware are rotations around arbitrary axes in the $X$-$Y$-plane, with 
propagator acting on qubit $j$~\cite{Pogorelov2021},
\begin{align}
	R_{j}\left(\theta,\varphi\right)&=
		\left[\begin{matrix}
		\mathrm{cos}\left(\theta/2\right) & -ie^{-i\varphi}\mathrm{sin}\left(\theta/2\right) \\
		 -ie^{i\varphi}\mathrm{sin}\left(\theta/2\right) & \mathrm{cos}\left(\theta/2\right)
		 \end{matrix}\right].
\end{align}
As demonstrated in~\cite{Arute2019} for superconducting qubits, this random ensemble of rotations can be configured to rapidly generate
entanglement when combined with deterministic entangling gates.
Inspired by this work, we fix the angle $\theta$ in the single-qubit rotations to $\theta=\pi/2$ and randomly choose one of three angles
$\varphi\in\left\{0,\pi/4,\pi/2\right\}$.
These gates describe $\pi/2$-rotations around the $X$-axis ($\varphi=0$), around the $Y$-axis ($\varphi=\pi/2$), and around the axis
between $X$ and $Y$ ($\varphi=\pi/4$).
We note that, though the MS gate and the rotations around the $X$- and $Y$-axis are Clifford gates, the single-qubit rotation with
$\left(\theta=\pi/2,\varphi=\pi/4\right)$ defines a gate beyond the Clifford group.
This choice makes it so that our circuit can not be simulated efficiently (i.e., in polynomial time) 
with a classical computer.  However, the local configuration of gates is amenable to tensor network simulations, which will
form the core of our numerical strategy below.

The combined application of single- and two-qubit gates with fixed angle choices can be summarized in a unitary 
operator:
\begin{align}
U_{i}\left(\varphi_i,\varphi_{i+1}\right)&=R_{i}\left(\pi/2,\varphi_{i}\right)
R_{i+1}\left(\pi/2,\varphi_{i+1}\right)M_{i,i+1}\left(\pi/4\right).
\end{align}
We thus build the brick-layer construction in the quantum circuit by alternately applying operators $U^{\mathrm{o}}$ and 
$U^{\mathrm{e}}$, with
\begin{align}
U^{\mathrm{o}}&=\prod_{i\ \mathrm{odd}}U_{i}\left(\varphi_{i},\varphi_{i+1}\right),\\
U^{\mathrm{e}}&=\prod_{i\ \mathrm{even}}U_{i}\left(\varphi_{i},\varphi_{i+1}\right),
\end{align}
and refer to odd and even unitary layers, respectively.
The random angles $\varphi_i$, $\varphi_{i+1}$ are chosen independently for each unitary operator 
$U_{i}\left(\varphi_i,\varphi_{i+1}\right)$.

By repeated application of $U^{\mathrm{o}}$ and $U^{\mathrm{e}}$, we create a circuit with $D$ layers of gates, each
evolving a pure quantum state $\ket{\Psi}$ as
\begin{align}
	\ket{\Psi}\rightarrow U^{k}\ket{\Psi}, \ k\in\left\{\mathrm{e},\mathrm{o}\right\}.
	\label{eq:unitary_dynamics}
\end{align}
We further define a time cycle as the combined application of odd and even unitary layers, so that a periodic behavior is 
induced in time and the circuit consists of $T=D/2$ time cycles~\cite{Li2019}, where one such cycle is depicted in \Fig{1}
for a system with $N=6$ qubits.

For large times $t$, the pure quantum state evolving under the dynamics in \Eq{unitary_dynamics} reaches a constant entanglement
entropy, which shows a volume-law behavior as we will demonstrate numerically in the next section.
In order to induce a phase transition into area-law entanglement at long times, we inject measurements uniformly through the circuit
(i.e.\ after each single MS layer) at a fixed rate $p$ (see \Fig{1}).
Specifically, at any space-time location, we perform a local projective measurement $P^{\pm}_i=\left(1_i\pm\sigma^z_i\right)/2$ into the
computational basis with a probability $p$, associated to the $\sigma^z$ Pauli matrix.
On average, $pND$ measurements are performed in the circuit, where each measurement projects the wavefunction according to the
measurement outcome $\sigma_i^z=\pm1$,
\begin{align}
	\ket{\Psi}\rightarrow \frac{P^{\pm}_i\ket{\Psi}}{\left\|P^{\pm}_i\ket{\Psi}\right\|}.
\end{align}
Independent of the results, these projections disentangle qubit $i$ from the rest of the system, and stand in contrast to the entangling
unitary gates.

\section{Tensor network implementation}
\label{sect:3}

The hybrid circuit that we propose in the last section is not amenable to any asymptotically efficient classical simulation scheme
[i.e., scaling $\mathrm{poly}(N)$ in time].
This is in contrast to Clifford circuits, for which efficient simulation is ensured through the Gottesman-Knill theorem~\cite{Gottesman1996,Gottesman1998,Gottesman2004}, or 
quantum automaton circuits, which are tractable despite their high entanglement~\cite{Iaconis2020,Iaconis2021}.
In order to simulate the long-time dynamics of our model circuit, we employ tensor-network methods for approximate time evolution.
Despite not being $\mathrm{poly}(N)$ scalable in general for the physics we explore, we will demonstrate below 
that these techniques allow us to reach circuit sizes sufficiently large to resolve
the universal properties of the MPT critical point.
Because of the design choice of circuit topology, we leverage efficient and well-refined algorithms based on matrix product states (MPS)~\cite{Garcia2007,Vestraete2008,Orus2014},
a class of parametric wavefunctions that has revolutionized large-scale simulations on strongly-correlated quantum matter in one
(and quasi-one) dimension~\cite{White1992,White1993,Schollwoeck2011,Stoudenmire2012}.

An MPS is a classical representation of a many-body wavefunction $\ket{\Psi}$, obtained by decomposing $\ket{\Psi}$ into a set of local tensors
$A_j^{\sigma_j}$:
\begin{align}
	\ket{\Psi}=\sum_{\bm\sigma}\prod_j A_j^{\sigma_j}\ket{\sigma_1,\dots,\sigma_N}\:.
\end{align}
Each rank-3 tensor (rank-2 at boundaries) has a physical index $\sigma_j$ and two auxiliary indices with 
dimension $\mathcal{D}_j$.
The total number of parameters scales as $O(N\mathcal{D}^2)$, where $\mathcal{D}=\mathrm{max}\{\mathcal{D}_j\}$ is the 
{\it bond dimension} of the MPS.
In the limiting case of $\mathcal{D}=1$ one recovers mean-field theory, while larger bond dimensions allow the MPS to capture
increasingly large amounts of quantum correlations.
This however comes at a cost that may become exponential in the number of qubits.

We evolve the initial product state using a procedure analogous to the time-evolving block decimation (TEBD) algorithm~\cite{Vidal2003,Vidal2004}.
We examine the MPS wavefunction after $T$ time cycles for different measurement rates $p$.
At each application of a gate, the contraction is approximated by only keeping a reduced number of states (i.e., the bond dimension)
following singular value decomposition, or by discarding all singular values below a fixed accuracy threshold $\varepsilon$.
Because of the orthogonality property of MPSs, this approximation remains controlled, in that the local truncation error is equal to the global error in approximating
the wavefunction~\cite{Schollwoeck2011,Cirac2020}. For $p>0$, independent measurements are realized by sampling the measurement outcome from the local one-body reduced density 
matrix, and projecting the MPS accordingly.

The numerical experiments shown in the following sections are performed using state-of-the-art Julia implementations of tensor network algorithms
for quantum circuit simulations.
We specifically use PastaQ.jl~\cite{pastaq}, a toolkit of tensor-network algorithms for simulation and benchmarking of quantum hardware,
based on the ITensors.jl package for efficient tensor manipulation~\cite{itensor}.
We tested our tensor network simulations for small circuit sizes against exact time evolution using the Yao.jl package~\cite{Yao}.

Before we turn to numerical calculations of physical properties in the next section, 
we first examine the simulation complexity associated with the tensor network method.
A natural measure of complexity is the bond dimension discussed above, which determines the total number of parameters in an MPS wavefunction.
Physically, the bond dimension indicates the amount of entanglement in the quantum state.
For the separable state $\ket{\Psi_0}$ at time $t=0$, the bond dimension required is $\mathcal{D}=1$.
During the random circuit dynamics, the bond dimension of the simulation grows to capture the spreading of correlations at long distances 
associated with increasing entanglement.
Because MPS wavefunctions possess exponentially decaying correlations, $\mathcal{D}$ may in fact grow very rapidly in time, 
hindering simulation in the low-$p$ regime.

\begin{figure}
	\includegraphics[width=\linewidth]{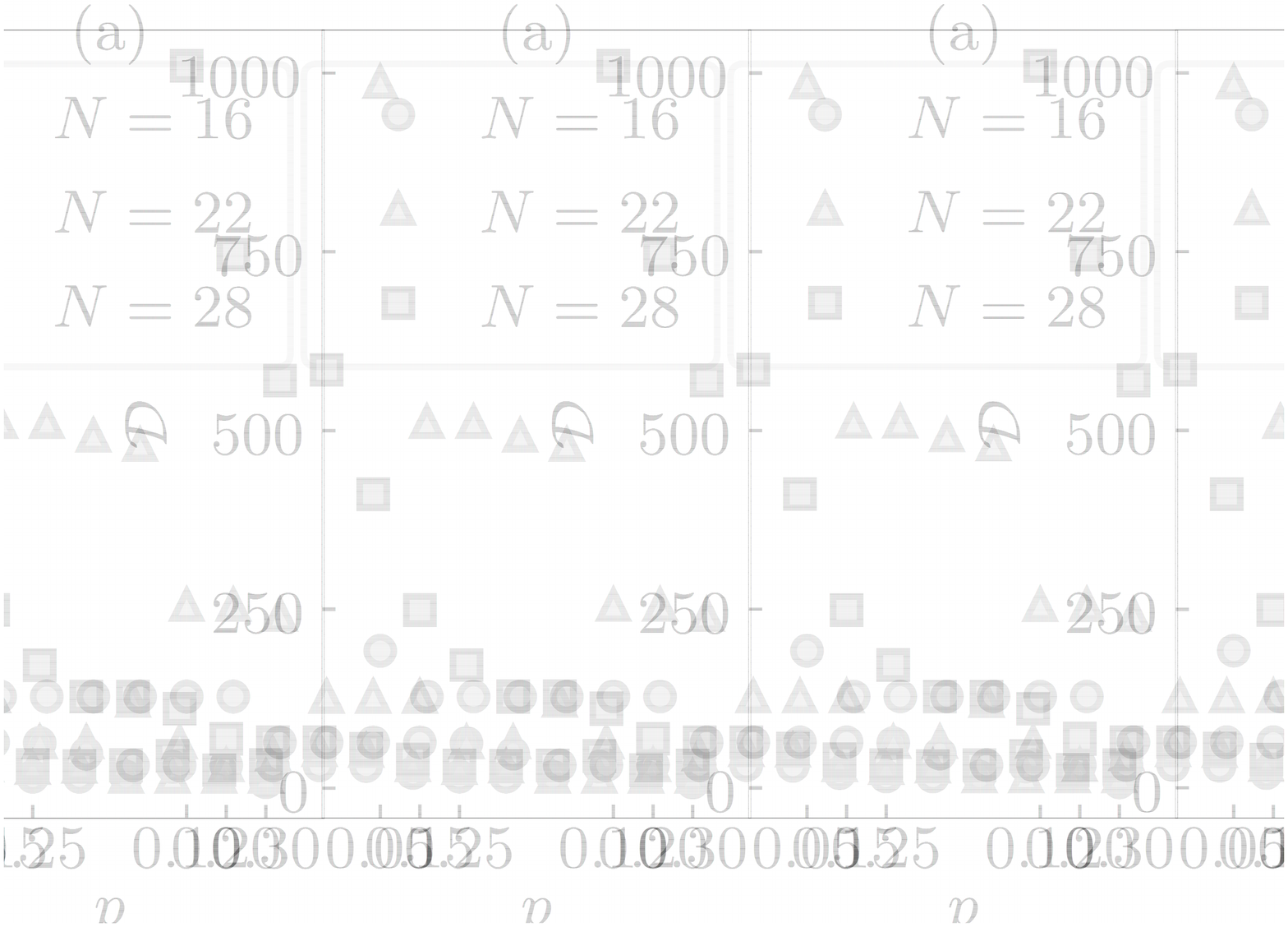}
	\caption{
	(a) The maximum bond dimension $\mathcal{D}$ observed in the long-time states of $500$ random circuits as a function of error rate $p$
	for tensor network simulations with a truncation cutoff of $\varepsilon=10^{-10}$.
	Different system sizes $N$ are considered, where for $N=28$ no data exists for $p\leq0.09$ due to exceeding computation times 
	for large bond dimensions.
	(b) Illustration of the entanglement phase diagram.
	A measurement-induced phase transition separates the two phases at the critical measurement rate $\pc$; see
	\Sect{4}.
	}
	\label{fig:2}
\end{figure}

We examine the computational cost required to reach quantum states deep in the volume-law phase (i.e., low $p$).
In \Fig{2} we show the maximum bond dimension reached in the long-time states of $500$ random circuits for different rates $p$ and increasing system size $N$, 
obtained with a fixed truncation cutoff of $\varepsilon=10^{-10}$.
For large measurement rates we find convergence to a constant value of $\mathcal{D}$, which does not depend on the system size.
This matches our expectations for an area-law entangling phase for large $p$.
In contrast, for small measurement rates we observe increasing bond dimensions with increasing $N$.
This is consistent with volume-law entanglement, and becomes more pronounced as $p$ decreases.
From these first observations, we can make a preliminary conclusion that a critical measurement rate $\pc$ may exist between $p\approx 0.17$ and $p\approx0.2$,
below which the dependence on the system size starts to form, as indicated in \Fig{2}(b).

\section{Simulation results}
\label{sect:4}

In order to characterize the putative phase transition, we focus on the entanglement entropy, 
the most relevant physical indicator of an MPT.
We focus on generalized $\alpha$-R{\'e}nyi entanglement entropies
\begin{align}
	S_{\alpha}\left(A\right)&=\frac{1}{1-\alpha}  \mathrm{ln}\left(\vphantom{\int}\mathrm{Tr}\left[\rho_{A}^\alpha\right]\right), 
	\label{eq:Renyi}
\end{align}
evaluated on subsystem $A$ after a bipartition of the state $\ket{\Psi}$ into subsystems $A$ and $\bar{A}$,
with reduced density matrix
\begin{align}
	\rho_{A}&=\mathrm{Tr}_{\bar{A}}\ket{\Psi}\bra{\Psi}.
\end{align}
In \Eq{Renyi} the limit $\alpha=1$ recovers the familiar von-Neumann entanglement entropy.
The second R{\'e}nyi entropy, $\alpha=2$, is most amenable to experimental measurement protocols~\cite{Islam2015,Brydges2019}.
In the tensor network simulations, the full family of R{\'e}nyi entropies is accessible from a Schmidt decomposition of the MPS at any 
given bond.

\subsection{Entanglement scaling}

In this section we detail our numerical results for the entanglement entropy of our hybrid circuit shown in \Fig{1}, where we
choose even numbers of qubits $N$ and open boundary conditions in accordance with numerical and experimental constraints.
Thus, the first and last qubits in the one-dimensional chain are not affected by MS gates in every second layer of unitaries (see \Fig{1}).
There are three sources of randomness that play a role in the dynamics: the choice of single-qubit rotations, the location of 
projective measurements, and the measurement outcomes~\cite{Li2019}.
We carry out independent simulations and average over these three realizations of disorder.
Specifically, we study entanglement entropies of half-chain bipartitions, so that the two subsystems contain equal numbers of qubits,
$\left|A\right|=N_A=N_{\bar{A}}=N/2$.
For convenience we write $S_{\alpha}\left(N,p,t\right)$, which refers to the half-chain entanglement entropy for a system of
$N$ qubits after $t$ time cycles in a circuit with measurement rate $p$.

\begin{figure}
	\includegraphics[width=\linewidth]{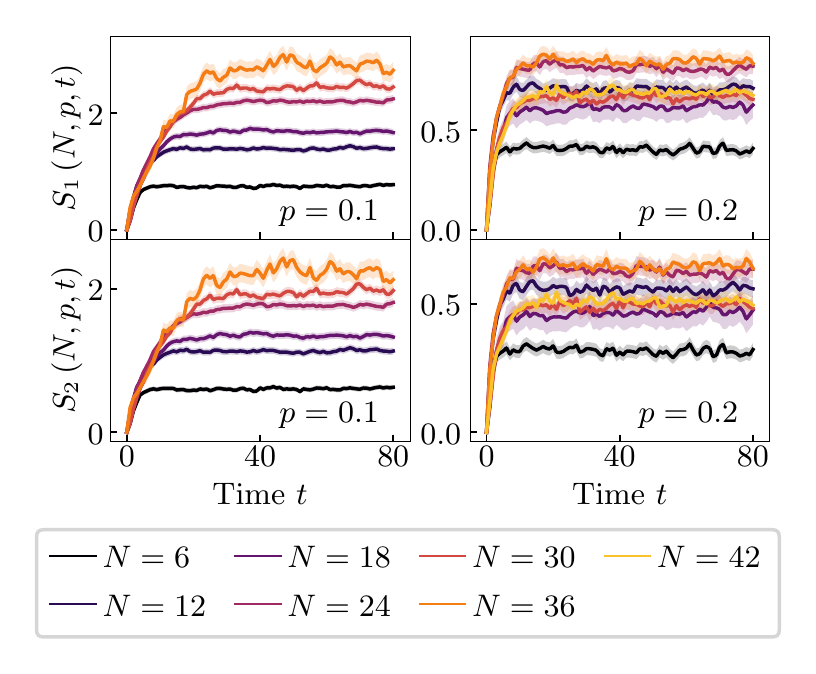}
	\caption{
	The evolution of the first and second half-chain entanglement R{\'e}nyi entropies under the quantum circuit are plotted as functions
	of time cycles $t$ for different system sizes $N$ and measurement rates $p=0.1$ (left) and $p=0.2$ (right).
	Convergence to a constant value for large $t$ indicates a saturated long-time entanglement entropy.
	The dependence on $N$ for $p=0.1$ for both measures indicates a volume-law entanglement, while the right column
	shows area-law entanglement since the entropies converge to similar values independent of the system size.
	All data points result from averaging $500$ circuit runs, except for sizes $N=30$ and $N=36$ with measurement rate 
	$p=0.1$, where only $85$ and $19$ circuit runs are averaged, respectively, as the large entanglement entropy 
	leads to excessive computation times.
	Shaded areas show the standard mean error, which is barely visible when averaging $500$ runs.
	Entanglement entropies for subsystems with an odd number $N_A=N/2$ of spins are smaller than those for 
	subsystems with even $N_A$ due to open boundary conditions~\cite{Laflorencie2006, Calabrese2009}.
	}
	\label{fig:3}
\end{figure}

We start with considering the evolution of the half-chain entanglement entropy under time 
cycles of the circuit for different system sizes $N$ and measurement rates $p$.
\Fig{3} shows $S_1\left(N,p,t\right)$ and $S_2\left(N,p,t\right)$ as functions of time cycles $t$, evaluated after each even unitary layer.
We consider two measurement rates $p=0.1$ and $p=0.2$ for different system sizes $N$, averaging over $500$ circuit runs.
All simulations show a clear convergence to a fixed entanglement entropy for $t\gtrsim N$ time cycles, 
confirming a circuit depth of $T=2N$ to study the constant long-time entanglement behavior.

While for $p=0.2$ the entanglement entropy saturates at similar values for all system sizes, a strong growth with increasing 
$N$ is observed for $p=0.1$.
This dependence on the subsystem size indicates volume-law entanglement for small measurement rates, while its absence 
for larger measurement rates is a sign for convergence to area-law entanglement at late times.
\Fig{3} hence suggests a phase transition between volume-law and area-law entanglement at a critical measurement rate 
$0.1\leq \pc\leq 0.2$.

As already observed in \Fig{2}, the simulations in the volume-law phase require large bond dimensions and hence excessive
computation times.
This limits our numerical studies with $p=0.1$ in \Fig{3} to smaller system sizes, so that for $N=30$ and $N=36$ only $85$ and $19$ circuit
runs are averaged, respectively, explaining stronger fluctuations.
Further, simulations are not feasible for $N>36$.

As an upper bound to the second-order R\'{e}nyi entropy~\cite{Chan2019}, the von-Neumann entropy takes larger values, 
while both entanglement measures show similar dependencies on system size and measurement rate.
This confirms that both entanglement measures can be used to detect the entanglement phase transition, which is also 
expected to happen at similar measurement rates~\cite{Li2019,Skinner2019,Jian2020,Zabalo2020}.
This further motivates an experimental realization of the quantum circuit to study MPTs.

\subsection{Critical behavior} \label{sect:CriticalSec}

\begin{figure}
	\includegraphics[width=\linewidth]{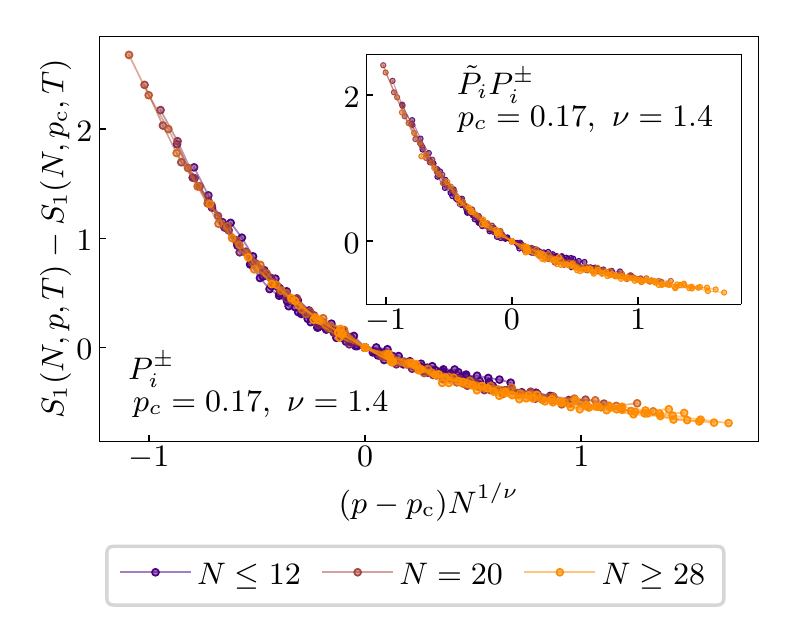}
	\caption{
	{\it Main plot}: The scaling behavior of the half-chain von-Neumann entanglement entropy [\Eq{4}] 
	for different system sizes $N\in\left\{6,8,\dots,36\right\}$, for circuits with projective measurements $P^{\pm}_i$.
	The critical measurement density $\pc$ and scaling exponent $\nu$ are adjusted to produce the best collapse, 	
	resulting in $\pc=0.17\pm0.02$ and $\nu=1.4\pm0.2$.
	All data points result from averaging the entanglement entropy after $T=2N$ time cycles of $500$ circuit runs and
        error bars are smaller than data points.
	{\it Inset}: The scaling behavior for additional applications of $\tilde{P}_i=\ket{0}\bra{0}+\ket{0}\bra{1}$
        after each measurement, with $N\in\left\{6,8,\dots,36\right\}$.
	An optimal collapse gives $\pc=0.17\pm0.02$, $\nu=1.4\pm0.2$ in this case; 
	see \Sect{GenMeasSec}.
	}
	\label{fig:4}
\end{figure}
In order to study the transition between the volume-law and the area-law phase in more detail and determine the critical 
measurement rate, we consider the scaling behavior of the von-Neumann entanglement entropy.
The constant entanglement entropy after $T=2N$ time cycles is expected to show the finite-size scaling form 
\cite{Li2019,Skinner2019,Li2020,Zabalo2020}
\begin{align}
	\SvN{N,p,T}=&\alpha\left(\pc\right)\mathrm{ln}\left[N\right]+F\left[\left(p-\pc\right)N^{1/\nu}\right],
	\label{eq:1}
\end{align}
with a factor $\alpha\left(\pc\right)$, scaling function $F$, and critical exponent $\nu$.
At criticality, we furthermore expect \cite{Li2019,Skinner2019,Zabalo2020}
\begin{align}
	\SvN{N,\pc,T}=&\alpha\left(\pc\right)\mathrm{ln}\left[N\right]+b,
	\label{eq:2}
\end{align}
with a constant offset $b$, and hence $F\left(0\right)=\mathrm{const.}$
We can determine the critical measurement rate $\pc$ and the critical exponent $\nu$ by eliminating the logarithmic 
term of the scaling form,
\begin{align}
	\SvN{N,p,T}-\SvN{N,\pc,T}=&\tilde{F}\left[\left(p-\pc\right)N^{1/\nu}\right],
	\label{eq:4}
\end{align}
and comparing the expected behavior with numerical calculations.

We perform tensor network simulations for different measurement rates $p$ and system sizes $N$ and average the 
entanglement entropy after $T=2N$ time cycles over $500$ circuit runs with randomly chosen single-qubit rotations and 
measurement placements.
Due to the large entanglement entropies in the volume-law phase, we consider smaller system sizes for small measurement
rates.

\Fig{4} shows the resulting collapse of the entanglement entropies to the scaling form in \Eq{4}, where we find 
$\pc=0.17\pm0.02$ and $\nu=1.4\pm0.2$ by optimizing the collapse.
The clearly observable universal scaling behavior at criticality confirms the expectation of an entanglement phase transition.
The determined critical exponent agrees with observations for different circuits in related works 
\cite{Li2019,Sang2020,Lavasani2021,Gullans2020,Gullans2020b,Zabalo2020,Shtanko2020}.

\begin{figure}
	\includegraphics[width=\linewidth]{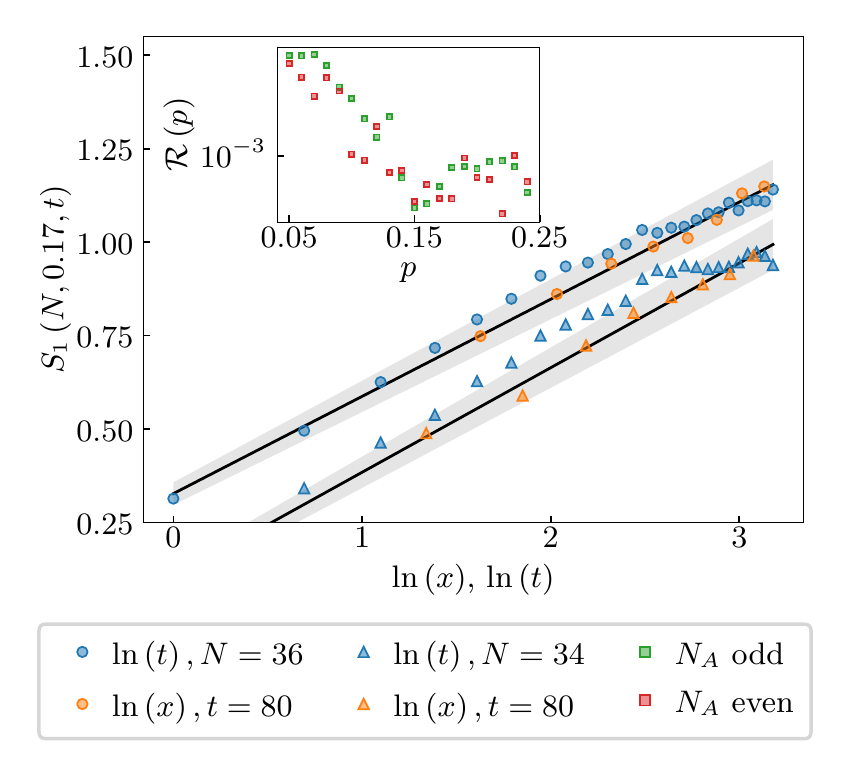}
	\caption{
	{\it Main plot}:
	Scaling function of the entanglement entropy close to the critical measurement rate $\pc$.
	$\SvN{N,0.17,t}$ is plotted as function of $\mathrm{ln}(t)$ for fixed system sizes $N=34$ (triangles) and $N=36$ (circles),
	and as a function of $\mathrm{ln}(x)$, $x=2N/\pi$, for system sizes $N\in\{6,8,\dots,36\}$ after $t=80$ time cycles,
	distinguishing even (circles) and odd (triangles) system sizes due to open boundary conditions~\cite{Laflorencie2006,Calabrese2009}.
	All data points are averaged over $500$ circuit runs and error bars are smaller than the data points.
	Black lines show functions $g(N)=\alpha(0.17)\mathrm{ln}(x)+\tilde{b}$ fitted to the data points, giving fit parameters $\alpha(0.17)=0.26\pm0.01$,
	$\tilde{b}=0.33\pm0.03$ for even and $\alpha(0.17)=0.28\pm0.01$, $\tilde{b}=0.1\pm0.03$ for odd subsystem sizes.
	Shaded regions show fit uncertainties.
	{\it Inset}: 
	Mean-squared error, \Eq{5}, of fitted scaling forms according to \Eq{2} versus measurement rate $p$ for odd and
	even subsystem sizes.
	A clear minimum around $p=0.15$ to $p=0.17$ indicates the critical measurement rate.
	Same data as in \Fig{4}.
	}
	\label{fig:5}
\end{figure}
Apart from the scaling collapse, we analyze the scaling form at criticality in \Eq{2}.
We estimate the critical measurement rate $\pc$ by fitting the scaling form to the numerical data at different measurement
rates $p$ and considering the fit quality, which is maximized at $\pc$.
The resulting data are shown in the inset of \Fig{5}, where we quantify the fit quality via the mean-squared error,
\begin{align}
	\mathcal{R}\left(p\right)=\left<\left[\SvN{N,p,T}-f_p\left(N\right)\right]^2\right>_{N},
	\label{eq:5}
\end{align}
with fitted function $f_p(N)=\alpha(p_c)\mathrm{ln}[N]+b$.
The average is taken over the data for different system sizes $N$, and we distinguish even and odd subsystem sizes, since
odd subsystem sizes show smaller entanglement entropies due to the open boundary conditions~\cite{Laflorencie2006,Calabrese2009}.
A clear minimum is observed around $p=0.15$ to $p=0.17$, which agrees well with the previously observed critical measurement
rate and confirms the MPT in this regime.

The main plot in \Fig{5} shows that a similar logarithmic scaling form is observed in the dynamical behavior of the entanglement
entropy in the critical regime for fixed system sizes.
We plot $\SvN{N,0.17,t}$ as a function of $\mathrm{ln}(t)$ for fixed system sizes $N=34$, $N=36$ to represent odd and even
subsystem sizes.
Additionally, we plot the entanglement entropy as a function of $\mathrm{ln}(2N/\pi)$ for different subsystem sizes 
$N_{\mathrm{A}}=N/2$ after $t=80$ time cycles, where we include an offset in the scaling form which is expected from 
conformal field theories (CFTs)~\cite{Calabrese2009, Iaconis2020,Li2020}.

We observe that both scaling forms, in time and in system size, behave similarly when independently considering odd and even 
subsystem sizes.
We further fit functions $g(N)=\alpha(\pc)\mathrm{ln}(2N/\pi)+\tilde{b}$ and $g(t)=\alpha_t(\pc)\mathrm{ln}(t)+\tilde{b}_t$ to the corresponding
data, where we take the uncertainty of $\pc=0.17\pm0.02$ into account and consider measurement rates in this regime.
We observe similar coefficients $\alpha(\pc)=0.3\pm0.15$ and $\alpha_t(\pc)=0.3\pm0.15$ for both even and odd subsystem sizes.
The ratio of the coefficients, $z=\alpha_t(\pc)/\alpha(\pc)=1.04\pm0.04$ describes a dynamical critical exponent whose value indicates
space-time conformal invariance~\cite{Calabrese2009,Iaconis2020,Li2020,Zabalo2020,Gullans2020}.
Thus, our simulations demonstrate the emergence of a CFT at the studied MPT.

\section{Further experimental considerations}

In the above, we have provided numerical evidence for an MPT in a model of a hybrid circuit using native trapped ion gates.
Despite the experimental feasibility of the MS and single-qubit gate set~\cite{Monroe2021}, as well as recent demonstrations of
high-precision optical addressing that enables measurements to be performed throughout the circuit~\cite{Shih2021}, several
considerations remain before such an MPT might be observed in real trapped ion hardware.
In this section, we address what we believe are the major experimental challenges that remain, with a particular focus on a
possible implementation in realistic trapped ion experiments.

\subsection{Measurement corrections from states outside the Hilbert space} \label{sect:GenMeasSec}
Projective measurements of trapped ion qubit states often populate states that are outside of the qubit Hilbert space. 
Such processes do not necessarily reduce the fidelity of state measurements, as the leakage onto unwanted states occurs
only for one of the two qubit states. 
For example, during the spin-dependent fluorescent measurement in $^{171}\rm{Yb}^+$ ions,
the state $\ket{F=1,m_F=0}$, representing the qubit state $\ket{1}$, leaks to $\ket{F=1,m_F=\pm 1}$ states, while the state $\ket{F=0,m_F=0}$,
representing the qubit state $\ket{0}$, is preserved \cite{Olmschenk2007}.
Re-initializing the ion in the projected qubit state may be experimentally demanding, requiring real-time decision
logic or transferring population via other meta-stable atomic states \cite{Myerson2008}.

A potential experimental implementation can hence be simplified by applying an additional operator $\tilde{P}_i=\ket{0}\bra{0}+\ket{0}\bra{1}$
after each projective measurement operator $P_i^{\pm}$.
Such an operation is routinely implemented with high fidelity in trapped ion experiments, using optical pumping techniques.
Any measured qubit then continues to evolve from the $\ket{0}$ state, which avoids the outcome-dependent
operation of bringing the ion onto the $\ket{1}$ state.
In the numerical simulations the wavefunction transforms according to
\begin{align}
  \ket{\Psi}\rightarrow\frac{\tilde{P}_iP^{\pm}_i\ket{\Psi}}{\left\|\tilde{P}_iP^{\pm}_i\ket{\Psi}\right\|},
\end{align}
which is implemented via the same probabilistic process as discussed earlier, followed by the application of $\tilde{P}_i$.

A comparison of numerical simulations using these additional operations after measurements with the previously discussed results 
(see \Sect{CriticalSec})
demonstrates that the MPT is not strongly influenced by this replacement.
The inset in \Fig{4} shows the resulting scaling collapse according to \Eq{4}, whose optimization gives $\pc=0.17\pm0.02$ 
and $\nu=1.4\pm0.2$, which agrees with our observations for projective measurements.
This result demonstrates a flexibility in adapting the measurement procedure to experimental restrictions and provides an important
step towards efficient experimental realizations of MPTs.

\subsection{Measurement crosstalk and SPAM errors}\label{sect:crosstalk}
The realization of constant measurement rates in trapped ion hybrid circuits can be possible using recent advances in single-ion
optical addressing~\cite{Shih2021,Crain2014}.
However, such an implementation comes with a source of error that directly affects the critical measurement rate.
Due to the micron-scale spatial separation of the trapped ions, the addressing of a single ion can leak onto direct neighbors
(crosstalk error), resulting in unwanted measurement of the neighboring qubit.
This process increases the probability with which each ion is measured and thus effectively increases the measurement rate,
shifting the critical measurement rate $\pc$ to smaller values.
Specifically, if a single ion is measured with probability $p$ at a measurement layer in the circuit, and for each measurement the
crosstalk error causes the additional measurement of an adjacent ion with probability $p_{\mathrm{d}}$, the effective measurement
rate can be considered as
\begin{align}
	\left(1+p_{\mathrm{d}}-pp_{\mathrm{d}}\right)p.
	\label{eq:7}
\end{align}
Here, the last term takes into account that an additional measurement caused by crosstalk errors has no effect if the ion is already
measured.
We further assume that left and right neighbors are addressed with equal probabilities and that the crosstalk probability is small,
so that terms of higher order in $p_{\mathrm{d}}$, resulting from crosstalk errors on both neighboring ions, are neglected.

In state-of-the-art trapped ion addressing schemes, the relative intensity cross-talk error
can be suppressed below a level of $10^{-4}$ at the neighboring ion~\cite{Shih2021,Crain2014}.
The state measurement time (spin-dependent fluorescence) itself can be limited to $10 \;\mathrm{\mu s}$ using high resolution imaging systems~\cite{Noek2013}.
Thus, the probability of an accidental measurement of the neighboring qubits ($p_\mathrm{d}$) can be kept at a few percent level. 

For example, if the state detection laser beam is at the saturation intensity of the $^2S_{1/2}-{^2}P_{1/2}$ transition on the target $^{171}\rm{Yb}^+$ qubit,
the probability of accidental measurement of the neighboring qubit via spontaneous emission from $10^{-4}$ level cross-talk can be kept at
$p_\mathrm{d}\approx 10^{-4}\times (2 \pi\times 20\;\rm{MHz}/6)\times 10 \;\mathrm{\mu s}\approx 0.02$~\cite{Berkeland2002,Noek2013}.
Here, $2\pi\times 20\rm{MHz}$ is the natural linewidth of the $^2P_{1/2}$ state used in the measurement process. Accidental measurement of neighboring qubits can also be mitigated by using separate atomic species (or separate isotopes) \cite{Inlek2017} as
processing qubits and read-out qubits,
where the qubit state is transferred from a processing qubit to a read-out qubit before measurement.

However, a fundamental limit to the probability of accidental measurements is given by the absorption of photons emitted from an addressed ion
by neighboring ions.
The probability of this effect depends strongly on the spatial distance $d$ between two ions and scales as $\sim1/d^2$.
In a typical experiment of $^{171}\rm{Yb}^+$ ions, a spatial distancing of $d\approx 5\mathrm{\mu m}$ can be achieved, so that this error is limited to $\lesssim 0.5\%$, see \App{1} for details.
The error can be further reduced by orders of magnitude using e.g.\,quantum charge-coupled devices (QCCD) \cite{Kielpinski2002}.

With the estimate in \Eq{7} we expect a shift of $\pc$ which is much smaller than the uncertainty in its value resulting from
our computational methods.
We confirmed via numerical studies of circuits including additional measurements with a relatively large leakage rate of
$p_{\mathrm{d}}=0.1$ that the shift in $\pc$ is not detectable within numerical uncertainties, emphasizing the robustness of the model to experimental
errors.

Apart from the crosstalk errors in the single-qubit measurements, further imperfections appear when preparing the initial state, 
applying MS and single-qubit gates, or performing global measurements of the final state.
Such state-preparation and measurement (SPAM) errors are on the order of a few percent in existing trapped ion hardware~\cite{Monroe2021}, 
and can also influence the critical measurement rate.
However, since these errors are of the same order as measurement crosstalk errors, we expect the effects on the MPT to be
similarly small.

Even though a noticeable shift of the critical measurement rate can result from the combination of multiple small imperfections, we expect
the resulting shift of $\pc$ to be small enough so that the MPT remains at finite measurement rates, $\pc>0$.
Thus, the MPT is robust against experimental imperfections and motivates an experimental realization.

\subsection{Estimators and post-selection}
Beyond the above experimental considerations, an important difference between our tensor network simulations and
real-world experiments is the procedure by which the long-time quantum wavefunction is characterized.
The evaluation of observables in trapped ion hardware is based on a statistical average of measurement outcomes,
typically performed on the final quantum state.
Most relevant for characterizing an MPT, the second R{\'e}nyi entropy can be extracted from statistical correlations
between local randomized measurement outcomes~\cite{Brydges2019, Enk2012}. It can hence be evaluated from
final-state measurement averages, much like any conventional observable.
However, the probabilistic behavior of random circuit (monitored) measurements causes further limitations. This comes from the
usual requirement of repeated state preparation and final-state measurements (possibly in multiple bases) in order to reconstruct any observable.
In the case of hybrid circuits created by monitored measurements spread throughout the circuit, this necessity of repeated
state preparation and measurement adds an additional requirement on post-selection, which naively results in an exponential scaling overhead.
In order to prepare a specific quantum state multiple times, the outcomes of all monitored single-qubit measurements need
to be identical in the considered circuit runs.
The probabilistic projections in the circuit for $p>0$ thus require the post-selection of the final quantum states of multiple
circuit runs based on the binary measurement outcomes.
For a hybrid quantum circuit of depth $T$ acting on $N$ ions, on average $\sim 2^{2pNT}$ circuit runs are needed in order to prepare a
specific quantum state, causing the exponential overhead in circuit runs~\cite{Gullans2020,Nahum2021,Bao2020}.

While microsecond-scale gate and measurement times of the trapped ion hardware will allow experiments with $N\approx 10$ qubits with measurement rates up to $p\approx 0.05$
within reasonable run times, the exponential post-selection overhead will pose a challenge for larger systems and higher measurement rates.
The achievable measurement rates are hence much smaller than the numerically observed critical measurement rate where the MPT happens in the studied circuit.
Reducing the critical measurement rate $p_c$ based on alternative circuit setups built of native ion gates might be a promising approach to reveal MPTs in a larger system. 

Accidental measurements appearing in experimental circuit runs due to crosstalk errors as discussed in \Sect{crosstalk} cause further issues in the post-selection procedure.
Since the outcomes of these stochastically appearing measurements are not monitored in the experiment, two similarly prepared final states can differ even when post-selecting the monitored measurement outcomes.
This strengthens the necessity of minimizing crosstalk effects in trapped ion hardware, which can be achieved e.g.\,by incorporating coupling of ions as in QCCD architectures \cite{Kielpinski2002}.
We numerically estimate the average loss of fidelity due to stochastic crosstalk measurements compared to the ideally prepared state without leakage errors to be $\lesssim 10\%$ for relevant system sizes and the crosstalk rates discussed in \Sect{crosstalk}.
Understanding the exact effect of this fidelity loss on the sample-based evaluation of entanglement entropies is an interesting topic for future studies.

While this post-selection procedure presents a clear bottleneck for experimental realizations of MPTs, recent works suggest ways to
bypass this measurement overhead.
Such proposals use additional ancilla qubits which are initially maximally entangled with the circuit qubits as a reference to locally probe the MPT~\cite{Gullans2020,Lavasani2021,Noel2021}.
A local order parameter can be associated with the average entropy of the reference qubits, which quantifies the amount of information shared with the initial
state and is hence large in the volume-law phase and small in the area-law phase, undergoing the same transition as the entanglement entropy.
The introduction of such local order parameters allows for the detection of the MPT at the critical measurement rate while reducing the necessary amount of
global measurements and avoiding the exponential post-selection overhead.
Furthermore, the number of additional qubits can be kept as small as a single reference qubit, making the approach feasible for experimental realizations~\cite{Gullans2020, Noel2021}.
Apart from improvements of this reference qubit approach based on cross-entropy estimators to reduce the amount of experimental measurements~\cite{Bao2020},
other proposals to circumvent the post-selection process are based on mappings onto space-time duals, which describe unitary dynamics and allow for an 
efficient measurement of R{\'e}nyi entanglement entropies~\cite{Ippoliti2021}.
Such theoretically developed techniques propose experimentally feasible ways to overcome the post-selection bottleneck by only slightly increasing the number
of necessary qubits, and may enable realizations of MPTs in trapped ion hardware.


\section{Discussion and outlook}

In this paper we have introduced a hybrid circuit model consisting of native trapped ion gates that is a strong candidate for
realizing a measurement-induced phase transition (MPT).  In order to induce entanglement growth, the circuit uses a
combination of two-qubit M{\o}lmer-S{\o}rensen gates and random single-qubit rotations.
To compete with this entanglement growth, measurements are induced throughout the circuit at a constant rate $p$.  
Such measurements should be realizable via recent experimental advances in high-precision, single-ion optical addressing schemes~\cite{Shih2021,Crain2014}.

The native gates studied in this work go beyond the Clifford group and can hence not be simulated efficiently with a scalable classical
algorithm.  In order to study this hybrid circuit, we therefore employ state-of-the-art tensor network simulations, evolving a 
matrix product state wavefunction using a time-evolving block decimation algorithm.
These simulations demonstrate two clear entanglement regimes in the long-time behavior of the circuit, corresponding to 
volume-law entanglement for small $p$ and area-law entanglement for large $p$.
Further, our numerical results provide clear evidence for an MPT at a critical measurement density $\pc \approx 0.17$.

In the area-law regime, our tensor network simulations are able to converge typical hybrid circuits with up to 44 qubits
and more than 10,000 unitary gates.  Such sizes are sufficiently large to resolve the critical properties of the MPT in great detail.
Collapsing our data near $\pc$ to a critical scaling function reveals the universal exponent $\nu=1.4\pm0.2$, which is within 
numerical uncertainty of the value of $\nu=4/3$ given by first-passage percolation in two dimensions~\cite{Skinner2019,Zabalo2020,Sang2020,Lavasani2021,Shtanko2020,Li2020,Jian2020,Li2019}.
In addition, the entanglement entropy shows clear logarithmic scaling at $\pc$, which agrees with previous studies of MPTs
that show broad features of conventional conformal field theories~
\cite{Calabrese2009,Sang2020,Li2019,Skinner2019,Jian2020,Li2020,Zabalo2020,Gullans2020}.
The coefficients of the logarithmic fits in the space and time direction directly at $\pc$ reveal coefficients that are the same to within numerical errors, 
indicating the apparent emergence of conformal symmetry.  

Our numerical results suggest that modern trapped ion experiments with a reasonable number of qubits are well-equipped to realize 
such MPTs.  We have argued that even in the presence of realistic experimental errors, such as measurement crosstalk errors, the MPT is sufficiently robust
to be induced with native entangling gates.
We have further shown that the qubits can be brought into outcome-independent states after the projective measurements without affecting the MPT, which
suggests a decision-logic free experimental implementation.
However, we also find that without significant modification of the gate set architecture, experimental studies 
may be limited to measurement rates far in the volume-law phase. This is due to a post-selection process required
for evaluating physical observables, including the second R{\'e}nyi entropy.  A significant body of theoretical work is currently
motivated by finding solutions to overcome this limitation~\cite{Gullans2020,Nahum2021,Lavasani2021,Bao2020,Ippoliti2021,Noel2021}.

Despite these challenges, our results indicate that trapped ion quantum computers are compelling platforms to realize 
MPTs, and could therefore conceivably be among the first engineered quantum systems to realize this exciting new class of 
physical phenomena in nature.  In addition, we have emphasized the crucial role that state-of-the-art numerical simulations 
will play in advancing the theoretical and experimental frontiers of MPTs and related physics in the future.

\section*{Acknowledgments}

We thank T. Hsieh, C.-Y. Shih, A. Vogliano, D. McLaren, C. Senko, R. Luo, M. P. A. Fisher, M. Stoudenmire, J. Iaconis, Z. Bandic, P. Bridger, and C. Noel for critically important discussions.
The calculations in this work were enabled in part by support provided by SHARCNET and Compute Canada.
We acknowledge financial support from Canada First Research Excellence Fund (CFREF) through the Tranformative Quantum Technologies (TQT) program, Natural Sciences and
Engineering Research Council of Canada's Discovery program, and Institute for Quantum Computing. RM is also supported by a Canada Research Chair. RI is also supported by
an Early Research Award from the Government of Ontario, and Innovation, Science and Economic Development Canada (ISED). Research at Perimeter Institute is supported in
part by the Government of Canada through the Department of Innovation, Science and Economic Development Canada and by the Province of Ontario through the Ministry of
Economic Development, Job Creation and Trade.

\setcounter{equation}{0}
\setcounter{figure}{0}
\setcounter{table}{0}
\makeatletter
\renewcommand{\theequation}{A\arabic{equation}}
\renewcommand{\thefigure}{A\arabic{figure}}
\renewcommand{\theHequation}{\theequation}
\renewcommand{\theHfigure}{\thefigure}
\appendix
\section{Fundamental limitations of crosstalk errors} 
\label{app:1}
The probability of crosstalk errors induced by absorption of photons emitted from an addressed ion can be kept at a small level.
This probability depends strongly on the spatial distance between two ions and can be estimated via the photon absorption cross section $\sigma$ of 
the neighboring ion.
For the considered detection line of the $^2S_{1/2}-{^2}P_{1/2}$ transition, the cross section scales as $\sigma\approx \lambda^2/(2\pi)$ with the 
detection wavelength $\lambda=369.5 \;\mathrm{nm}$ for $^{171}\rm{Yb}^+$.
This corresponds to a cross section radius of $a=\sqrt{\sigma/\pi}\approx 80 \;\mathrm{nm}$.
Thus, a fraction proportional to $(1-\mathrm{cos}(\theta))/(8\pi)\approx 6.4\times 10^{-5}$ of the photons emitted by the targeted ion is absorbed, with the 
angle $\theta=\mathrm{arctan}(a/d)$ subtended by the photon absorption cross section area.
For an optimal scattering rate of $7\times 10^6 \;\mathrm{Hz}$ for the target $\rm{Yb}^+$ ion and detection times of $10 \;\mathrm{\mu s}$, this gives
a crosstalk error probability of $p_\mathrm{d}\approx 5\times 10^{-3}$.
This probability can be increased when applying the corrective measurement scheme discussed in \Sect{GenMeasSec}, since the additional 
pumping of an ion emits further photons.


%

\end{document}